\documentclass[12pt]{article}
\usepackage{amsfonts,amsmath,graphicx,setspace}
\usepackage{color,hyperref,verbatim,natbib,rotating}

\definecolor{Red}{rgb}{0.5,0,0}
\definecolor{Blue}{rgb}{0,0,0.5}
\hypersetup{%
    colorlinks = {true},
    linktocpage = {true},
    plainpages = {false},
    linkcolor = {Blue},
    citecolor = {Blue},
    urlcolor = {Red},
    pdfstartview = {XYZ null null 1.25},
    pdfpagemode = {UseOutlines},
    pdfview = {XYZ null null null}
}
\newcommand{\email}[1]{\href{mailto:#1}{\normalfont\texttt{#1}}}
\newcommand{\bfalpha}{\boldsymbol{\alpha}}
\newcommand{\bfbeta}{\boldsymbol{\beta}}
\newcommand{\bfpsi}{\boldsymbol{\psi}}
\newcommand{\bfgamma}{\boldsymbol{\gamma}}
\newcommand{\bftheta}{\boldsymbol{\theta}}
\newcommand{\bflambda}{\boldsymbol{\lambda}}
\newcommand{\bfdelta}{\boldsymbol{\delta}}
\newcommand{\bfDelta}{\boldsymbol{\Delta}}
\newcommand{\bfTheta}{\boldsymbol{\Theta}}

\newcommand{\by}{\mathbf{y}}
\newcommand{\bt}{\mathbf{t}}
\newcommand{\br}{\mathbf{r}}
\newcommand{\bs}{\mathbf{s}}
\newcommand{\bx}{\mathbf{x}}
\newcommand{\bz}{\mathbf{z}}
\newcommand{\bw}{\mathbf{w}}

\newcommand{\bb}{\mathbf{b}}
\newcommand{\bu}{\mathbf{u}}
\newcommand{\dd}{{\sf d}}

\newcommand{\bY}{\mathbf{Y}}
\newcommand{\bX}{\mathbf{X}}
\newcommand{\bT}{\mathbf{T}}
\newcommand{\bD}{\mathbf{D}}
\newcommand{\bK}{\mathbf{K}}
\newcommand{\bO}{\mathbf{O}}

\bibpunct{(}{)}{;}{a}{}{,}

\begin{document}

{\vspace*{1cm}}

\begin{center}
\Large \bf Goodness-of-Fit Checks for Joint Models
\end{center}

\vspace{0.5cm}

\begin{center}
{\large Dimitris Rizopoulos$^{1,2,*}$, Jeremy M.G. Taylor$^{3}$ and Isabella Kardys$^{4}$}\footnote{$^*$Correspondence at: Department of Biostatistics, Erasmus University Medical Center, PO Box 2040, 3000 CA Rotterdam, the Netherlands. E-mail address: \email{d.rizopoulos@erasmusmc.nl}.}\\
$^{1}$Department of Biostatistics, Erasmus University Medical Center, the Netherlands\\
$^{2}$Department of Epidemiology, Erasmus University Medical Center, the Netherlands\\
$^{3}$Department of Biostatistics, University of Michigan, Ann Arbor, USA\\
$^{4}$Department of Cardiology, Erasmus University Medical Center, the Netherlands\\
\end{center}

\begin{spacing}{1}
\noindent {\bf Abstract}\\
Joint models for longitudinal and time-to-event data are widely used in many disciplines. Nonetheless, existing model comparison criteria do not indicate whether a model adequately fits the data or which components may be misspecified. We introduce a Bayesian posterior predictive checks framework for assessing a joint model's fit to the longitudinal and survival processes and their association. The framework supports multiple settings, including existing subjects, new subjects with only covariates, dynamic prediction at intermediate follow-up times, and cross-validated assessment. For the longitudinal component, goodness-of-fit is assessed through the mean, variance, and correlation structure, while the survival component is evaluated using empirical cumulative distributions and probability integral transforms. The association between processes is examined using time-dependent concordance statistics. We apply these checks to the Bio-SHiFT heart failure study, and a simulation study demonstrates that they can identify model misspecification that standard information criteria fail to detect. The proposed methodology is implemented in the freely available \textsf{R} package \textbf{JMbayes2}.\\\\
\noindent {\it Keywords:} Dynamic Predictions, Longitudinal Data Analysis, Posterior Predictive Checks; Survival analysis; Time-varying covariates.
\end{spacing}

\vspace{0.8cm}

%=====================================================

\section{Introduction} \label{Sec:Introduction}
Joint models for longitudinal and time-to-event data have attracted significant attention in recent years \citep{rizopoulos:12, papageorgiou.et.al:19}. Several variants of these models have been proposed, and their implementation is now available in widely used statistical software. Yet, despite their growing complexity and widespread adoption, there is a lack of systematic tools for assessing goodness-of-fit. Such tools are of practical importance for joint models, as a common critique is that they rely on parametric assumptions that are difficult to evaluate.

Fitted joint models are often compared using the deviance information criterion (DIC) and Watanabe-Akaike information criterion (WAIC). \citet{zhang.et.al:11} have proposed a novel decomposition of the Akaike Information Criterion and Bayesian Information Criterion that allows for assessing the fit of each component of the joint model. While these metrics will suggest which model is best, they do not provide any reassurance that the model is actually fitting the data, nor do they offer any assessment of which aspects of the joint model are compatible with, or not so compatible with, the data. 

We introduce a comprehensive framework of posterior predictive checks for joint models that assesses the goodness-of-fit of the longitudinal and survival components and their association. We focus on three evaluation settings: (i) assessing the fit for existing subjects conditioning on their longitudinal, event time, and baseline covariate information, (ii) new subjects for whom we only have covariate information but no longitudinal measurements, and (iii) existing or new subjects with observed longitudinal histories, covariate information, and no event by that time. For the longitudinal outcome, we assess model fit for the mean and variance functions over time, and the correlation structure of the repeated measurements. For the event outcome, we evaluate the model fit using both marginal and individual level survival functions, and we use a concordance statistic to assess if the model adequately describes the association structure between the two processes.

Our work is motivated by the Bio-SHiFT study,\citep{brankovic.et.al:18} in which we are interested in studying the association between renal dysfunction and adverse outcomes in patients with chronic heart failure. In particular, we aim to investigate how the temporal evolutions of creatinine-estimated glomerular filtration rate (eGFR) and plasma neutrophil gelatinase-associated lipocalin (NGAL) are associated with the risk of the composite endpoint, including heart failure hospitalization, cardiac death, left ventricular assist device placement, and heart transplantation, using joint models.

The rest of the paper is organized as follows: Section~\ref{Sec:JMs} introduces the joint models' framework. Section~\ref{Sec:PPCs} presents the framework of posterior predictive checks for joint models. Section~\ref{Sec:GoF} presents the goodness-of-fit statistics for evaluating the model's fit. Section~\ref{Sec:BioSHiFT_Analysis} shows the results from the joint model we fitted in the Bio-SHiFT study, and Section~\ref{Sec:Simulation} shows the results of a simulation study that exemplifies the use of the proposed checks in a control setting. Finally, Section~\ref{Sec:Discussion} closes the paper with a discussion.

%=====================================================

\section{Joint Modeling Framework} \label{Sec:JMs}
\subsection{Joint Models} \label{Sec:JMs_def}
In this section, we present a general definition of the joint modeling framework for longitudinal and time-to-event data. Let $\mathcal D_n = \{T_i, \delta_i, \by_i, \mathcal X_i; i = 1, \ldots, n\}$ denote a sample from the target population, where $T_i^*$ denotes the true event time for the $i$-th subject, $C_i$ the censoring time, $T_i = \min(T_i^*, C_i)$ the corresponding observed event time, and $\delta_i = \mathbb{I}(T_i^* \leq C_i)$ the event indicator, with $\mathbb{I}(\cdot)$ being the indicator function that takes the value 1 when $T_i^* \leq C_i$, and 0 otherwise. In addition, we let $\by_i$ denote the $n_i \times 1$ continuous longitudinal response vector for the $i$-th subject, with element $y_{il}$ denoting the value of the longitudinal outcome taken at time point $t_{il}$, with $\{t_{il} < T_i; l = 1, \ldots, n_i\}$. Finally, $\mathcal X_i$ denotes a set of baseline covariates for the $i$-th subject.

We postulate that the response vector $\by_i$ conditional on the vector of unobserved random effects $\bb_i$ has a distribution $\mathcal F_\psi$ parameterized by the vector $\bfpsi$. This more general formulation allows for distributions that are not in the exponential family. The mean of the distribution of the longitudinal outcome conditional on the random effects has the form:
\[g \bigl [ E \{ y_i(t) \mid \bb_i \} \bigr ] = \eta_i(t) = \bx_i^\top(t) \bfbeta + \bz_i^\top(t) \bb_i,
\]
where $g(\cdot)$ denotes a known one-to-one monotonic link function, and $y_i(t)$ denotes the value of the longitudinal outcome for the $i$-th subject at time point $t$, $\bx_i(t)$ and $\bz_i(t)$ (both determined $\mathcal X_i$ and $t$) denote the time-dependent design vectors for the fixed-effects $\bfbeta$ and for the random effects $\bb_i$, respectively. We let $\sigma$ denote the scale parameter of $\mathcal F_\psi$, i.e., $\bfpsi^\top = (\bfbeta^\top, \sigma)$. The random effects are assumed to follow a multivariate normal distribution with mean zero and variance-covariance matrix $\bD$, i.e., $\bb_i \sim \mathcal{N}(\mathbf{0}, \bD)$. For the survival process, we assume that the risk of an event depends on a function of the subject-specific linear predictor $\eta_i(t)$ and the random effects. More specifically, we assume
\begin{eqnarray*}
h_i \{t, \mathcal H_i(t), \bw_i\} & = &
\lim_{s \rightarrow 0} \frac{1}{s}\Pr \{ t \leq T_i^* < t + s \mid T_i^* \geq t, \mathcal H_i(t),
\bw_i \}\\
& = & h_0(t) \exp \bigl  [\bfgamma^\top
\bw_i + f \{\mathcal H_i(t), \bb_i, \bfalpha \} \bigr] , \quad t > 0,
\end{eqnarray*}
where $\mathcal H_i(t) = \{ \eta_i(s), 0 \leq s < t \}$ denotes the history of the underlying longitudinal process up to $t$, $h_0(\cdot)$ denotes the baseline hazard function, $\bw_i$ is a vector of baseline covariates (a subset of $\mathcal X_i$)  with corresponding regression coefficients $\bfgamma$. The baseline hazard function $h_0(\cdot)$ is modeled flexibly using a B-splines approach, i.e.,
\[
\log \{ h_0(t) \} = \sum \limits_{p = 1}^P \gamma_{h_0,p} B_p\{r(t), \bflambda\},
\]
where $B_p\{r(t), \bflambda\}$ denotes the $p$-th basis function of a B-spline (that includes an intercept) with knots $\lambda_1, \ldots, \lambda_P$ and $\bfgamma_{h_0}$ the vector of spline coefficients. The function $r(\cdot)$ denotes a potential transformation of the time variable. For example, selecting $r(t) = \log(t)$ and restricting the B-spline basis to a linear fit correspond to the Weibull hazard function.

The function $f(\cdot)$, parameterized by vector $\bfalpha$, specifies the components of the underlying longitudinal outcome process that are included in the linear predictor of the relative risk model. Some examples are:
\begin{eqnarray*}
f \{\mathcal H_i(t), \bb_i, \bfalpha \} = \left \{
\begin{array}{l}
\alpha \eta_i(t),\\
\alpha \eta_i'(t),
\mbox{ with } \eta_i'(t) = \frac{\dd \eta_i(t)}{\dd t},\\
\alpha \eta_i''(t),
\mbox{ with } \eta_i''(t) = \frac{\dd^2\eta_i(t)}{\dd t^2},\\
\alpha \displaystyle \frac{1}{v} \displaystyle \int_{t - v}^t \eta_i(s) \, \dd s, \;\; 0 < v \leq t,\\
\alpha \displaystyle \frac{1}{v} \displaystyle \int_{t - v}^t \eta_i''(s) \, \dd s, \;\; 0 < v \leq t.
\end{array}
\right.
\end{eqnarray*}
These formulations of $f(\cdot)$ postulate that the hazard of an event at time $t$ is associated with the underlying level of the biomarker at the same time point, the slope/velocity of the biomarker at $t$, the acceleration of the biomarker at $t$, the average biomarker level in the period $(t - v, t)$, or the curvature of the biomarker profile in the same period. Combinations of these functional forms and their interactions with baseline covariates are also often considered.

%%%%%%%%%%%%%%%%%%

\subsection{Bayesian Estimation} \label{Sec:JMs_estimation}
We let $\bftheta^\top = (\bfbeta^\top, \sigma, \bfgamma_{h_0}^\top, \bfgamma^\top, \bfalpha^\top, \mbox{vech}(\bD)^\top, \tau)$ denote the vector of all model parameters, where $\mbox{vech}(\bD)$ denotes the unique elements of the variance-covariance matrix $\bD$. We estimate the model under the Bayesian paradigm and draw inferences using the joint posterior distribution $\{\bftheta, \bb \mid \bY, \bT, \bfdelta, \mathcal X\}$, where $\bY$, $\bT$, $\bfdelta$, and $\mathcal X$ denote outcome and covariate vectors for all $n$ subjects. We use standard priors for $\bftheta$, i.e., normal priors for all regression coefficients $(\bfbeta, \bfgamma, \bfgamma_{h_0}, \bfalpha)$, inverse-Gamma priors for $\sigma$ and the diagonal elements of $\bD$, and the LKJ prior for the correlation matrix of the random effects. To ensure smoothness of the baseline hazard function $h_0(t)$, we postulate a `penalized' prior distribution for the regression coefficients $\bfgamma_{h_0}$ that is proportional to:
\[
\tau^{\rho(\bK)/2}\exp \Bigl (-\frac{\tau}{2} \bfgamma_{h_0}^\top \bK \bfgamma_{h_0} \Bigr ),
\]
where $\tau$ is the smoothing parameter that takes a $\mbox{Gamma}(5, 0.05)$ hyper-prior in order to ensure a proper posterior for $\bfgamma_{h_0}$, $\bK = \bfTheta_r^\top \bfTheta_r$, where $\bfTheta_r$ denotes $r$-th difference penalty matrix, and $\rho(\bK)$ denotes the rank of $\bK$. When a simpler model is assumed for the baseline hazard (e.g., the Weibull hazard or regression splines instead of penalized splines) the $\tau$ parameter is excluded from $\bftheta$. We use a Markov chain Monte Carlo (MCMC) approach to obtain samples from the posterior distribution for all model parameters and the random effects. This algorithm is implemented in the freely available \textsf{R} package \textbf{JMbayes2} \citep{JMbayes2} that we used to fit the model.

%=====================================================

\section{Posterior Predictive Checks} \label{Sec:PPCs}
\subsection{Posterior-Posterior Predictive Checks} \label{Sec:PPC_Post-Post}
To investigate the fit of a joint model to the training dataset $\mathcal D_n$, we compare simulated longitudinal measurements $\by_i^{\sf rep}$ and event times $T_i^{\sf *rep}$ from the fitted model with the observed data $\mathcal D_n$ \citep{gelman.et.al:96, gelman.et.al:21}. Ideally, we expect that realizations from the fitted model are in close agreement with the observed data. The simulation of new realizations from the fitted joint model should also carefully consider the visiting and censoring processes \citep{tsiatis.davidian:04}. In particular, we let $\mathcal{V}_i(t) = \{t_{il}: 0 \leq t_{il} < t; l = 1, \ldots, n_i(t)\}$ denote the visit times at which the $i$-th subject provided longitudinal measurements up to follow-up time $t$, and $n_i(t)$ denote the number of these measurements. We also let $\mathcal{Y}_i(t) = \{y_i(t_{il}); l = 1, \ldots, n_i(t)\}$ denote the observed longitudinal measurements before $t$. Finally, we let $\mathcal{T}_i(t) = \{\mathbb{I}(T_i^* < s); 0 \leq s < t\}$ denote the counting process for an event occurring before time $t$, and $\mathcal{C}_i(t) = \{\mathbb{I}(C_i < s); 0 \leq s < t\}$ denote if the $i$-th subject was censored before $t$. We write the joint distribution of these processes, including also the current longitudinal measurement $y_i(t)$:
\begin{eqnarray*}
\lefteqn{p(\{y_i(t), \mathcal{Y}_i(t)\}, \mathcal{T}_i(t), \mathcal{V}_i(t), \mathcal{C}_i(t) \mid \mathcal{X}_i)}\\
& = & p(\{y_i(t), \mathcal{Y}_i(t)\}, \mathcal{T}_i(t) \mid \mathcal{X}_i) \;\; p(\mathcal{V}_i(t), \mathcal{C}_i(t) \mid \{y_i(t), \mathcal{Y}_i(t)\}, \mathcal{T}_i(t), \mathcal{X}_i),
\end{eqnarray*}
where, $p(\cdot)$ denotes a probability density or a probability distribution function, accordingly. The standard assumption made by joint models is that the visiting and censoring processes only depend on previously observed longitudinal responses and covariates, but are independent of future longitudinal measurements and the random effects, i.e.,
\[p(\mathcal{V}_i(t), \mathcal{C}_i(t) \mid \{y_i(t), \mathcal{Y}_i(t)\}, \mathcal{T}_i(t), \mathcal{X}_i) =
p(\mathcal{V}_i(t) \mid \mathcal{Y}_i(t), \mathcal{X}_i) \;\; p(\mathcal{C}_i(t) \mid \mathcal{Y}_i(t), \mathcal{X}_i).
\]
Under this assumption, inference for the joint distribution of the parameters $\{\bftheta, \bb_i; i = 1, \ldots, n\}$ proceeds via the joint distribution of the outcomes $\prod_i p(\{y_i(t), \mathcal{Y}_i(t)\}, \mathcal{T}_i(t) \mid \mathcal{X}_i)$, while ignoring the visiting and censoring processes (we also assume that the parameters $\{\bftheta, \bb_i; i = 1, \ldots, n\}$ are functionally independent from the parameters of the visiting and censoring processes).

Under the same assumption, the posterior predictive distribution for subject $i$ based on a joint model has the following general form:
\begin{eqnarray*}
p(\bO_i^{\sf rep} \mid T_i, \delta_i, \by_i, \mathcal{X}_i, \mathcal D_n) & = & \int\int p(\bO_i^{\sf rep} \mid \bb_i, \mathcal{X}_i, \bftheta) \, p(\bb_i \mid T_i, \delta_i, \by_i, \mathcal{X}_i, \bftheta) \\
&& \quad \quad \quad \quad \, p(\bftheta \mid \mathcal D_n) \, \dd \bb_i \dd \bftheta,
\end{eqnarray*}
where $\bO_i^{\sf rep}$ denotes either $\by_i^{\sf rep}$ or $T_i^{\sf *rep}$, or both. In the second term, we make explicit that the random effects vector $\bb_i$ only depends on the data of subject $i$. Hence, the simulation of replicated data from the model entails the following steps:
\begin{enumerate}
\item Sample a value $\{\tilde{\bftheta}, \tilde{\bb}_i\}$ from the MCMC sample of the posterior distribution $[\bftheta, \bb_i \mid T_i, \delta_i, \by_i, \mathcal{X}_i, \mathcal D_n]$.

\item Sample a value $\by_i^{\sf rep}$ from $\mathcal{F}(\tilde{\bfpsi}, \tilde{\bb}_i, \mathcal{X}_i)$ or $T_i^{\sf *rep}$ from $\mathcal{S}( \tilde{\bfgamma}_{h_0}, \tilde{\tau}, \tilde{\bfgamma}, \tilde{\bfalpha}, \tilde{\bb}_i, \mathcal{X}_i)$, with $\mathcal{S}$ denoting the distribution of the time-to-event outcome defined from the hazard function presented above.
\end{enumerate}
Given the ignorability assumption, we can, in principle, simulate longitudinal measurements $\by_i^{\sf rep}$ at any time points $\{t_{il}^{\sf rep}; i = 1, \ldots, n; l = 1, \ldots, n_i\}$. However, the choice of the statistic we will use to compare the simulated data with the observed data, may restrict the manner we will simulate the replicated data. We will expand on this issue in the next section. Also, note that because we use the random effects that condition on the observed longitudinal responses, the observed event time, the event indicator and covariates, we can simulate $\by_i^{\sf rep}$ independently of $T_i^{\sf *rep}$.

%%%%%%%%%%%%%%%%%%

\subsection{Posterior-Prior Predictive Checks} \label{Sec:PPC_Post-Prior}
We term the predictive checks presented above as `posterior-posterior predictive checks' because we sampled the random effects $\bb_i$ for the posterior distribution given the observed data of subject $i$, that is, $\{\by_i, T_i, \delta_i, \mathcal{X}_i\}$. Alternatively, we can perform posterior predictive checks for generic subjects from our target population (i.e., subjects for whom we have no longitudinal or event time information, but we do have covariate information) by simulating data from the distribution \citep{gelman.et.al:96}:
\begin{eqnarray*}
p(\bO_i^{\sf rep} \mid \mathcal{X}_i, \mathcal D_n) & = & \int\int p(\bO_i^{\sf rep} \mid \bb_i, \mathcal{X}_i, \bftheta) \, p(\bb_i \mid \bftheta) \, p(\bftheta \mid \mathcal D_n) \, \dd \bb_i \dd \bftheta,
\end{eqnarray*}
that is, the random effects $\bb_i$ are now simulated from their prior distribution. In this context, the simulation Step 1 above is split into two sub-steps:
\begin{enumerate}
\item Sampling of parameters
\begin{enumerate}
    \item Sample a value $\tilde{\bftheta}$ from the MCMC sample of the posterior distribution $[\bftheta \mid \mathcal D_n]$.
    \item Sample a value $\tilde{\bb}_i$ from the prior $\mathcal{N}(\mathbf{0}, \tilde{\bD})$.
\end{enumerate}
\end{enumerate}
We refer to these as `posterior-prior predictive checks'. The simulation of $\by_i^{\sf rep}$ and $T_i^{\sf *rep}$ is done jointly, using the same random effect values generated in Step 1b. However, because the random effects are now simulated from their prior distribution and do not condition on the event times, no information is shared between the event and longitudinal processes. To make the link between the two processes explicit, we exclude the longitudinal measurements collected for a subject after their event time. That is, the occurrence of an event causes informative dropout in the longitudinal outcome. This step was not required in the posterior-posterior checks because the random effects used to simulate the longitudinal responses were sampled from the conditional distribution $[\bb_i \mid T_i, \delta_i, \by_i, \mathcal{X}_i]$.

%%%%%%%%%%%%%%%%%%

\subsection{Dynamic-Posterior-Posterior Predictive Checks} \label{Sec:PPC_Dyn-Post-Post}
A popular use of joint models is the calculation of dynamic predictions for longitudinal and survival outcomes \citep{rizopoulos:11, taylor.et.al:13, rizopoulos.taylor:24}. The advantageous feature of these predictions is that they can be updated over time as additional information about the subject becomes available during follow-up. In this context, it is relevant to assess a joint model's fit at different follow-up times, conditioning on the available information. In particular, we assume that we collected longitudinal measurements $\by_i(t_L)$ up to time $t_L$, and for the individuals at risk at this time point, we are interested in assessing the fit after $t_L$. Having collected measurements up to $t_L$ also implies that $T_i^* > t_L$ or equivalently that $\{T_i = t_L, \delta_i = 0\}$. Thus, the posterior predictive distribution takes the form:
\begin{eqnarray*}
\lefteqn{p \{ \widetilde{\bO}_i^{\sf rep} \mid T_i = t_L, \delta_i = 0, \by_i(t_L), \mathcal{X}_i, \mathcal{D}_n\}}\\ & = & \int\int p(\widetilde{\bO}_i^{\sf rep} \mid \bb_i, \mathcal{X}_i, \bftheta) \, p \{ \bb_i \mid T_i = t_L, \delta_i = 0, \by_i(t_L), \mathcal{X}_i, \bftheta \} \\
&& \quad \quad \quad \quad \, p(\bftheta \mid \mathcal{D}_n) \, \dd \bb_i \dd \bftheta,
\end{eqnarray*}
where $\widetilde{\bO}_i^{\sf rep}$ denotes the longitudinal measurements after $t_L$ or the true event time given that the subject was event-free up to $t_L$. To sample from this distribution requires adjusting the sampling of the random effects presented in Step 1 above, i.e.,
\begin{enumerate}
\item Sampling of parameters
\begin{enumerate}
    \item Sample a value $\tilde{\bftheta}$ from the MCMC sample of the posterior distribution $[\bftheta \mid \mathcal D_n]$.
    \item Sample a value $\tilde{\bb}_i$ from the conditional distribution $[\bb_i \mid T_i = t_L, \delta_i = 0, \by_i(t_L), \mathcal{X}_i, \tilde{\bftheta}]$.
\end{enumerate}
\item Sample a value for $\widetilde{\bO}_i^{\sf rep}$ using $\{\tilde{\bftheta}, \tilde{\bb}_i\}$.
\end{enumerate}
The distribution in Step 1b does not have a closed form, so we sample from it using a Metropolis-Hastings algorithm with a normal proposal distribution. In particular, the target probability density function is proportional to
\[
p(\by_i(t_L) \mid \bb_i, \mathcal{X}_i, \tilde{\bftheta}) \; \exp \biggl ( - \int_0^{t_L} h_i \{s, \mathcal H_i(s), \bw_i, \tilde{\bfgamma}_{h_0}, \tilde{\bfgamma}, \tilde{\bfalpha}\} \dd s \biggr ) \; p(\bb_i \mid \tilde{\bftheta}).
\]

The simulation of replicated data $\widetilde{\bO}_i^{\sf rep}$ in Step 2 also needs to be adjusted compared to the same step in Section~\ref{Sec:PPC_Post-Post}. In particular, for $\by_i^{\sf rep}$ we sample from $\mathcal{F}(\tilde{\bfpsi}, \tilde{\bb}_i, \mathcal{X}_i)$ for $t > t_L$, and we sample $T_i^{\sf *rep}$ from the conditional cumulative distribution function,
\[
F(t; t > t_L) = 1 - \exp \biggl ( - \int_{t_L}^t h_i \{s, \mathcal H_i(s), \bw_i, \tilde{\bfgamma}_{h_0}, \tilde{\bfgamma}, \tilde{\bfalpha}\} \dd s \biggr ).
\]

%%%%%%%%%%%%%%%%%%

\subsection{Cross-Validated Checks} \label{Sec:PPC_CV}
A criticism of standard posterior predictive checks is that the training dataset $\mathcal{D}_n$ is used both to fit and to evaluate the model's goodness-of-fit, leading to potentially over-optimistic results \citep{moran.et.al:24}. To alleviate such concerns, we propose to evaluate the model fit in an independent test dataset. In the absence of an external test dataset, we can utilize a cross-validation procedure. In particular, we split the original dataset $\mathcal{D}_n$ into $V$ folds. We denote by $\mathcal{D}_n^{(v)}$ the data in the $v$-th fold, and $\mathcal{D}_n^{(-v)}$ the dataset that excludes the measurements of the $v$-th fold. For a generic subject $j$ from $\mathcal{D}_n^{(v)}$, we compare their observed outcome data $\by_j$ and $\{T_j, \delta_j\}$ with replicated outcomes after fitting the model to $\mathcal{D}_n^{(-v)}$, i.e., $[\by_j^{\sf rep} \mid \mathcal{D}_n^{(-v)}]$ and event times $[T_j^{\sf *rep} \mid \mathcal{D}_n^{(-v)}]$. In particular, the posterior-prior predictive distribution is of the form:
\begin{eqnarray*}
p(\bO_j^{\sf rep} \mid \mathcal{X}_j, \mathcal{D}_n^{(-v)}) & = & \int\int p(\bO_j^{\sf rep} \mid \bb_j, \mathcal{X}_j, \bftheta) \, p(\bb_j \mid \bftheta) \, p(\bftheta \mid \mathcal{D}_n^{(-v)}) \, \dd \bb_j \dd \bftheta.
\end{eqnarray*}
Analogously, the predictive distribution for the (dynamic) posterior-posterior predictive checks takes the form:
\begin{eqnarray*}
\lefteqn{p \{\bO_j^{\sf rep} \mid T_j = t_L, \delta_j = 0, \by_j(t_L), \mathcal{X}_j, \mathcal{D}_n^{(-v)}\} }\\
& = & \int\int p(\bO_j^{\sf rep} \mid \bb_j, \mathcal{X}_j, \bftheta) \, p \{ \bb_j \mid T_j = t_L, \delta_j = 0, \by_j(t_L), \mathcal{X}_j, \bftheta \} \\
&& \quad \quad \quad \quad \, p(\bftheta \mid \mathcal{D}_n^{(-v)}) \, \dd \bb_i \dd \bftheta.
\end{eqnarray*}
For the dynamic posterior-posterior checks, $t_L$ is set equal to the observed time $T_j$ and $\delta_j$ is set to 1 if this subject had an event at this time. Also, in this case, and contrary to the posterior-posterior predictive checks for the subjects in the fitted dataset presented above, we need to sample the random effects using the Metropolis-Hastings scheme introduced in the dynamic posterior-posterior predictive checks. This is because we do not obtain a sample from the conditional distribution $[\bb_j \mid T_j, \delta_j, \by_j, \mathcal{X}_j, \bftheta]$ from the model fitted in $\mathcal{D}_n^{(-v)}$.

%=====================================================

\section{Goodness-of-Fit Measures} \label{Sec:GoF}
\subsection{Longitudinal Outcome} \label{Sec:GoF_Long}
A standard fit measure used in the context of posterior predictive checks is to compare the distribution of the observed outcome with the distributions of the simulated outcomes is the empirical cumulative distribution function (eCDF). Following this recommendation, we use the eCDF as a global fit measure to evaluate the fit of the joint model to the marginal distribution of the longitudinal outcome.

However, the eCDF does not provide insight into key aspects of the longitudinal outcome distribution, namely, the mean, variance and correlation structure. In particular, the mean function describes the average longitudinal evolution of the outcome, the variance function the longitudinal outcome's variance over time, and the correlation structure the pairwise correlations of the longitudinal responses as a function of the time lags (i.e., time elapsed) between measurements. To evaluate the joint model's fit to the mean function of the longitudinal submodel, we compare the local polynomial regression curve (loess) curve over time of the observed data $\by_i$ with the loess curves of the replicated outcomes $\by_i^{\sf rep}$. We do this comparison both for individual subjects and for the data pooled across all subjects.

To estimate the variance function of the longitudinal outcome, we need first to subtract its systematic trend over time. Typically, this is done by regressing $\by_i$ on the fixed-effects design matrix $\bX_i$ with ordinary least squares, and calculating the residuals
\[
\br_i^{\sf OLS} = \by_i - \bX_i \widehat{\bfbeta}_i^{\sf OLS}.
\]
However, when we wish to compare the fit of models with different fixed-effect parts, this results in different variance functions based on the same observed data $\mathcal{D}_n$. To alleviate the dependence of the variance function on the selection of the fixed-effects design matrix, we calculate the non-parametric residuals from the loess curve for the observed data $\by_i$ and the follow-up times $\bt_i$ (i.e., the estimated mean function presented above). The residuals are then defined as
\[
\br_i = \by_i - \widehat{\by}_i,
\]
with $\widehat{\by}_i$ denoting the estimated loess curve. We do the same computation for the replicated data and obtain $\br_i^{\sf rep}$. We then calculate the standardized residuals $\br_{si} = \br_i / \hat{\sigma}$, with $\sigma = \sqrt{\sum_i \br_i^\top \br_i / (N - {\sf df}(\widehat{\by}_i))}$, with $N = \sum_i n_i$ and ${\sf df}(\widehat{\by}_i)$ denoting the degrees of freedom of the loess curve. Likewise, we calculate the standardized residuals for the replicated data $\br_{si}^{\sf rep} = \br_i^{\sf rep} / \hat{\sigma}^{\sf rep}$. To depict the magnitude of the residuals' scale we calculate the transformations $\breve{\br}_{si} = \sqrt{|\br_{si}|}$ and $\breve{\br}_{si}^{\sf rep} = \sqrt{|\br_{si}^{\sf rep}|}$, respectively. Finally, we compare the loess curve over time of the observed data transformed residuals $\breve{\br}_{si}$ with the loess curves of the replicated transformed residuals $\breve{\br}_{si}$.

To investigate if the assumed random-effects structure captures the correlations in the longitudinal measurements sufficiently well, we use the sample semi-variogram. For a stochastic process $y(t)$, and a time lag $v$, the semi-variogram is defined as \citep{diggle.et.al:02}:
\[
\varphi(v) = \frac{1}{2} E \bigl [\{y(t) - y(t - v)\}^2 \bigr ], \quad v \geq 0,
\]
and when $y(t)$ is stationary, the semi-variogram is related to the auto-correlation function $\rho(v)$ via the formula:
\[
\varphi(v) = \sigma^2 \{1 - \rho(v)\},
\]
where $\sigma^2$ is the variance of $y(t)$. To estimate $\varphi(v)$, we start by removing the systematic trend from the longitudinal outcome using the loess residuals $\br_i$ and $\br_i^{\sf rep}$, as defined above for the variance function. Then, per subject, we calculate the half squared differences and the time lags
\begin{eqnarray*}
\Delta_{i,jk} = 0.5 (r_{ij} - r_{ik})^2 \quad \mbox{and} \quad u_{i,jk} = |t_{ij} - t_{ik}|, \;\; \mbox{with}\;\; \{j,k = 1, \ldots, n_{i}; j \neq k\}.
\end{eqnarray*}
We compare the loess curve over the time lags $u_{i,jk}$ of the semi-variogram $\Delta_{i,jk}$ for the observed residuals with the loess curves of the semi-variogram for the replicated residuals $\Delta_{i,jk}^{\sf rep}$. The variance and correlation structure functions presented above are typically calculated using the data from all subjects. However, in principle they could also be used to investigate the fit of the model for individual subjects.

We return to the topic of selecting the time points $\{t_{il}^{\sf rep}; i = 1, \ldots, n; l = 1, \ldots, n_i\}$ at which we simulate replicated data $\by_i^{\sf rep}$ from the joint model. The statistics presented above do not account for a visiting process that depends on previously observed longitudinal responses and covariates. Hence, if the visiting process depends on the history, but we simulate at times $t_{il}^{\sf rep}$ that are completely random, we may observe differences between the simulated and observed data statistics attributed to the differences in the visiting processes. To alleviate such discrepancies, we simulate the longitudinal measurements $\by_i^{\sf rep}$ at the same time points as the original measurements were collected $\{t_{il} < T_i; l = 1, \ldots, n_i\}$.

%%%%%%%%%%%%%%%%%%

\subsection{Time-to-Event Outcome} \label{Sec:GoF_Surv}
We simulate true event times $T_i^{\sf *rep}$ from the posterior predictive distribution of the joint model, ignoring the censoring process (i.e., we do not simulate censoring times). As we have done for the longitudinal outcome, we use the eCDF of these event times as a global fit measure. However, to compare the eCDF of the simulated data with the observed data, we need to account for censoring in the observed data $\{T_i, \delta_i; i = 1, \ldots, n\}$. If censoring is independent of the observed data, we use the Kaplan-Meier estimate of the cumulative distribution function. If censoring depends on covariates, we can estimate the eCDF using the Breslow estimator from a Cox model that includes these covariates.

 A second fit measure that focuses on individual level survival functions, is to use the probability integral transform of the subject-specific cumulative distribution functions. In particular, we calculate the eCDF of each subject $\{u_i = F_i(t); i = 1, \ldots, n\}$ using their simulated data $\{T_{im}^{\sf *rep}; m = 1, \ldots, M\}$, with $m$ denoting the realization of the simulated event times for the $i$-th subject, i.e.,
\[
u_i = \frac{1}{M} \sum\limits_{m = 1}^M \mathbb{I}(T_{im}^{\sf *rep} \leq t).
\]
Using the pairs $\{u_i, \delta_i\}$, where $\delta_i$ is the event indicator, we calculate the Kaplan-Meier estimate of the probability $\widehat{F}_i = \Pr(U_i \leq u)$. If the joint model fits the marginal event time distribution well, we expect the $\widehat{F}_i$ to be the cumulative distribution function of the uniform distribution. This approach also assumes that censoring is independent of observed data.

%%%%%%%%%%%%%%%%%%

\subsection{Association between Longitudinal and Event Time Outcomes} \label{Sec:GoF_Ass}
To evaluate whether the postulated joint model adequately describes the association between the longitudinal and event-time processes, we use the concordance statistic, which measures agreement between the time-to-event outcome and the longitudinal predictor. In particular, for each unique event time $\{t_0 = 0, t_1, \ldots, t_K\}$, we select the subjects at risk at $t_k$ and we denote by $\ddot{y}_i(t_k)$ their last available longitudinal measurement before $t_k$. We then calculate the concordance statistic $C(t_k)$ using the subjects at risk at $t_k$ and the last values $\ddot{y}_i(t_k)$.

This procedure is performed for the observed and the replicated longitudinal and event time data, yielding the pairs of variables $\{t_k, C(t_k); k = 0, \ldots, K\}$ and $\{t_k, C^{\sf rep}(t_k); k = 0, \ldots, K\}$. As we did above, we compare the loess curve of the observed concordance statistic over time with the loess curves of the concordance statistic from the replicated data. In the calculation of the concordance statistic, we have assumed a monotonic relationship between $\ddot{y}_i(t_k)$ and the event times. This assumption could be relaxed by calculating $C(t_k)$ from a Cox model for the subjects at risk at $t_k$ and including a nonlinear effect of $\ddot{y}_i(t_k)$ using splines or polynomials. Moreover, because the observed event times $T_i$ are subject to censoring, whereas the replicated times are not, commonly used C-statistics will be inappropriate because they depend on the study-specific censoring distribution. Hence, to appropriately account for censoring, we are using the procedure proposed by \citet{uno.et.al:11} that consistently estimates $C(t_k)$, which is free of censoring.

When the latest longitudinal measurement $\ddot{y}_i(t_k)$ is a long time before this risk set time, it is questionable whether such a measurement should be included in the calculation of the concordance statistic. A simple ad hoc adjustment for this type of situation is to discard longitudinal observations in the calculation of $C(t_k)$ that are, say, more than $\kappa$ time units before the risk time $t_k$.

%%%%%%%%%%%%%%%%%%

\subsection{Numerical Summary Measure} \label{Sec:GoF_MISE}
As a summary measure for the goodness-of-fit figures described above, we calculate the mean integrated squared error
\[
{\sf MISE} = \frac{1}{M}\sum_{m = 1}^M \int \bigl \{ F(\bs_m^{\sf rep}, \tilde{\bt}) - F(\bs, \tilde{\bt}) \bigr \}^2 \; \dd\tilde{\bt},
\]
where $F(\bs, \tilde{\bt})$ denotes the empirical cumulative distribution function of $\bs$ or the loess curve calculated from the pairs $\{\bs, \tilde{\bt}\}$, $\bs$ denotes either the outcome data $\by$, the standardized residuals $\br_s$, the half-squared-differences $\bfDelta$ or the concordance statistic $C(t_k)$. Analogously, $\bs_m^{\sf rep}$ ($m = 1, \ldots, M$) denotes the $m$-th realization of the replicated data $\bs^{\sf rep}$, and $\tilde{\bt}$ denotes the time points or the time lags $\bu$, respectively. The integral is approximated using the trapezoidal rule over the range of the observed data, with 200 sub-intervals.

%=====================================================

\section{Bio-SHiFT Data Analysis} \label{Sec:BioSHiFT_Analysis}
Renal dysfunction is an important factor in chronic heart failure (CHF). Using data from the Bio-SHiFT study, which included patients with CHF, we aim to investigate how the temporal evolution of creatinine-estimated glomerular filtration rate (eGFR) and plasma neutrophil gelatinase-associated lipocalin (NGAL) is associated with the risk of the composite endpoint, including heart failure hospitalization, cardiac death, left ventricular assist device placement, and heart transplantation. Patients were included in the study if aged $\geq 18$ years, capable of understanding and signing informed consent, and if CHF had been diagnosed $\geq 3$ months ago according to European Society of Cardiology guidelines. From the 263 patients enrolled in the first inclusion round of the study, 70 experienced the primary endpoint. The median (IQR) for the number of both eGFR and NGAL measurements was 9 (5-10). Section 1.1 in the supplementary material shows descriptive figures for the two longitudinal outcomes, and the Kaplan-Meier estimate for the composite event.

For each longitudinal outcome, we postulated two linear mixed models, one with linear and one with nonlinear subject-specific evolutions. In both, a set of baseline covariates has been included. In particular, the linear models have the form:
\begin{eqnarray*}
y_i(t) & = & (\beta_0 + b_{i0}) + (\beta_1 + b_{i1}) t \; + \\
&& \quad \tilde{\beta_1}\texttt{Age}_i + \tilde{\beta_2}\texttt{Male}_i + \tilde{\beta_3}\texttt{NYHA}_i + \tilde{\beta_4}\texttt{DM}_i + \tilde{\beta_5}\texttt{IHD}_i + \tilde{\beta_6}\texttt{Diuretics}_i +
\varepsilon_i(t),
\end{eqnarray*}
and analogously the nonlinear ones
\begin{eqnarray*}
y_i(t) & = & (\beta_0 + b_{i0}) + (\beta_1 + b_{i1}) NS_1(t) + (\beta_2 + b_{i2}) NS_2(t) + (\beta_3 + b_{i3}) NS_3(t)\; + \\
&& \quad \tilde{\beta_1}\texttt{Age}_i + \tilde{\beta_2}\texttt{Male}_i + \tilde{\beta_3}\texttt{NYHA}_i + \tilde{\beta_4}\texttt{DM}_i + \tilde{\beta_5}\texttt{IHD}_i + \tilde{\beta_6}\texttt{Diuretics}_i +
\varepsilon_i(t),
\end{eqnarray*}
where $y_i(t)$ denotes either the eGFR or the log2-transformed NGAL, $\{NS_k(t); k = 1, 2, 3\}$ denotes the basis of a natural cubic spline with three degrees of freedom, $\texttt{Age}_i$ the age of patient $i$, $\texttt{Male}_i$ is the dummy for males, $\texttt{NYHA}_i$ denotes the New York Heart Association class, $\texttt{DM}_i$ is the dummy for diabetes mellitus, $\texttt{IHD}_i$ is the dummy for ischemic heart disease, and $\texttt{Diuretics}_i$ is the dummy for diuretics use. The random effects $\bb_i$ follow a normal distribution with mean zero and variance-covariance matrix $\bD$, and the error terms have a normal distribution with mean zero and variance $\sigma^2$. Even though we use the same notation in the model definition for eGFR and NGAL, the parameters are different in the two models.

For the composite endpoint we postulated two relative risk models, namely
\begin{eqnarray*}
h_i(t) & = & h_0(t) \exp \{ \gamma_1\texttt{Age}_i + \gamma_2\texttt{Male}_i +
\gamma_3\texttt{NYHA}_i + \gamma_4\texttt{DM}_i + \gamma_5\texttt{IHD}_i +
\gamma_6\texttt{Diuretics}_i \; +\\
&& \quad \quad \quad \alpha_1 \eta_i^{\sf eGFR}(t) + \alpha_2 \eta_i^{\sf NGAL}(t) \},
\end{eqnarray*}
and
\begin{eqnarray*}
h_i(t) & = & h_0(t) \exp \Bigl \{ \gamma_1\texttt{Age}_i + \gamma_2\texttt{Male}_i +
\gamma_3\texttt{NYHA}_i + \gamma_4\texttt{DM}_i + \gamma_5\texttt{IHD}_i +
\gamma_6\texttt{Diuretics}_i \; + \\
&& \quad \quad \quad \alpha_1 \frac{1}{t} \int_0^t \eta_i^{\sf eGFR}(s) \; {\sf d}s +
\alpha_2 \frac{1}{t}\int_0^t \eta_i^{\sf NGAL}(s) \; {\sf d}s \Bigr \},
\end{eqnarray*}
The first model postulates that the current values of eGFR and NGAL are associated with the risk of the composite event at the current time. In contrast, the second model assumes that the average of eGFR and NGAL from entry to the study to time $t$ is associated with the risk of the composite event at $t$. Time $t = 0$ is the study inclusion time. We fitted the four bivariate joint models obtained by considering the possible combinations of mixed and relative risk models. Section 1.2 in the supplementary material shows parameter estimates and 95\% credible intervals for the four joint models. We observe that NGAL is more strongly associated with the risk of composite event compared to eGFR. The Deviance Information Criterion (DIC), the Watanabe-Akaike information criterion (WAIC), and the log pseudo marginal likelihood (LPML) do not uniformly agree in which of the four models provides the best fit/predictions to the data.

\begin{figure}
\centering{\includegraphics[width=\textwidth]{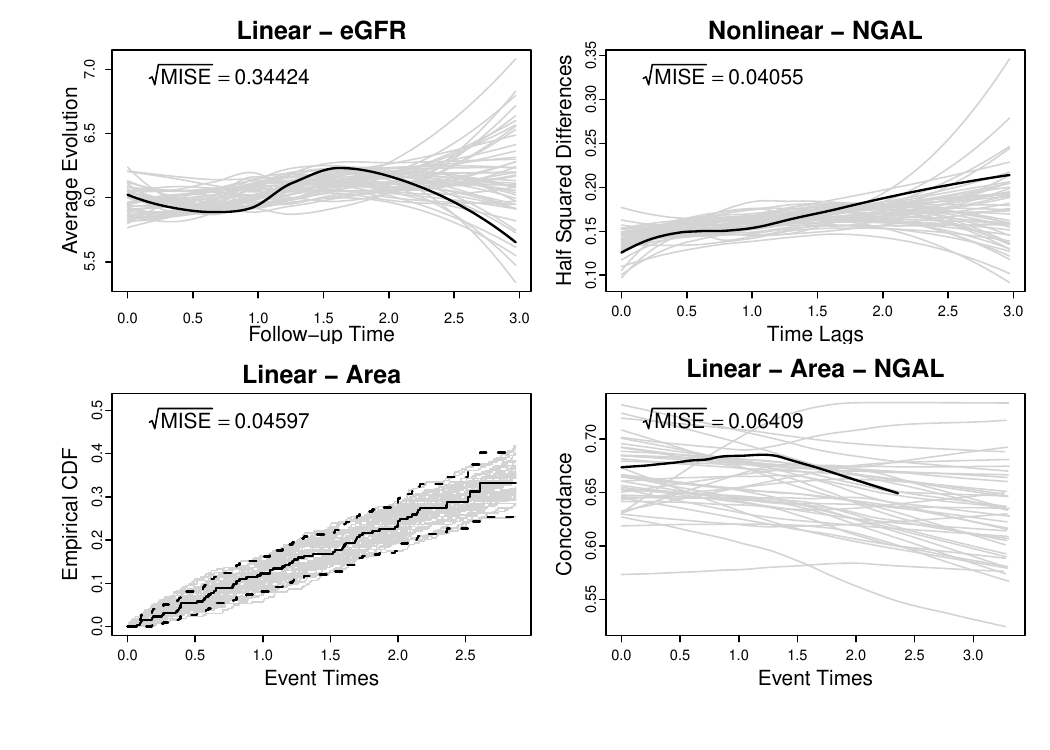}}
\caption{Top-left panel: posterior-posterior predictive checks for the mean function of eGFR using 50 simulated datasets from the joint model that assumes linear subject-specific profiles and the value functional form; the grey lines are the loess curves of the simulated data, and the superimposed black line is the loess curve of the observed data. Top-right panel: posterior-posterior predictive checks for the semi-variogram function of NGAL using 50 simulated datasets from the joint model that assumes nonlinear subject-specific profiles and the value functional form; the grey lines are the loess curves of the simulated data, and the superimposed black line is the loess curve of the observed data. Bottom-left panel: posterior-prior predictive checks for the composite event outcome using 50 simulated datasets from the joint model that assumes linear subject-specific profiles and the area functional form; the grey lines are the eCDF curves of the simulated data, the superimposed black line is the Kaplan-Meier curve of the observed data, and the dashed black line is the 95\% confidence interval of the Kaplan-Meier estimate. Bottom-right panel: posterior-prior predictive checks for the concordance statistic between NGAL and the composite event using 50 simulated datasets from the joint model that assumes nonlinear subject-specific profiles and the area functional form; the grey lines are the loess curves of the simulated data, and the superimposed black line is the loess curve of the observed data.} \label{Fig:sample-plots}
\end{figure}

We evaluate the fit of these models in the Bio-SHiFT database for both the longitudinal and event time outcomes using the posterior-posterior and posterior-prior predictive checks. We also used 10-fold cross-validation and fitted the same models using each time the subset of nine folds, and simulated replicated outcome data for the fold left outside. By combining the results from all folds, we evaluate the models' fit using posterior-prior and posterior-posterior checks. Using cross-validation, we also assess the fit for the patients at risk at $t_L = 1$ year. For the longitudinal outcome, we evaluate the global fit using the empirical cumulative distribution function and the first two moments using the mean, variance, and auto-correlation functions. For the event-time outcome, we use both the empirical cumulative distribution function and the probability-integral transformation. For the association between the two processes, we use the concordance statistic. The results are presented in detail Sections~1.3--1.10 of the supplementary material. As an example, we show in Figure~\ref{Fig:sample-plots} some of the posterior predictive checks we performed.

The eCDF plots for the two longitudinal outcomes suggest that the NGAL model has a better fit than the eGFR model. The specification of linear versus nonlinear trajectories does not appear to have a big influence on the eCDF plots. The mean function plots indicate that the models with nonlinear trajectories better describe the observed data, especially for the two patients whose data is presented. For generic patients from the same population, the differences are smaller. For the variance function, we observe minor differences between the linear and nonlinear models. The correlation structure in the repeated eGFR measurements seems to be somewhat better described by the model with linear subject-specific evolutions than the nonlinear one. The correlations in the NGAL repeated measurements seem to be adequately captured by both mean structures. In general, the fit of the longitudinal submodels seems better in the posterior-posterior than in the posterior-prior predictive checks. Also, the cross-validation assessment of the models' fit indicates that they perform less well on completely unseen data. However, when we condition on the longitudinal measurements recorded up to $t_L = 1$ year of follow-up (the median number of measurements before $t_L$ was 4 for both outcomes), the fits improve considerably. From the individualized checks for the mean functions of the two longitudinal outcomes, we observe that the data patterns of Patient~180 are less well fit by all four models, with the nonlinear models doing a bit better. On the contrary, the fit of all models to the data of Patient~124 is adequate. The checks for the survival submodel do not indicate any major issues for the fit of this part of the joint models to the event process. The checks for the association between the risk of the composite event and eGFR suggest that the area/integral functional form provides a better fit than the current value one. The association between the composite event and NGAL is well described under both functional forms. The values of the concordance statistic are higher for NGAL than they are for eGFR, which is compatible with the results for the parameter estimates and credible intervals.

%=====================================================

\section{Simulation Study} \label{Sec:Simulation}
To investigate the usefulness of the posterior predictive checks presented above in a controlled setting, we present the evaluation of a joint model's fit in a simulated dataset with 300 subjects. In particular, we simulated longitudinal responses from the linear mixed-effects model:
\[
\left \{
\begin{array}{l}
y_i(t_{il}) = \eta_i(t_{il}) + \varepsilon_i(t_{il}),\\
\eta_i(t_{il}) = (\beta_{0} + b_{0i}) + \sum \limits_{k=1}^3(\beta_{k} + b_{1k}) NS_k(t_{il}). \\
\end{array}
\right.
\]
The value for the fixed-effects coefficients were $(\beta_0, \ldots, \beta_3) = (1.75, 0.033, -5.84, -0.182)$, and $NS_k(\cdot)$ denote a natural cubic spline for the time variable, with internal knots placed at follow-up times $\{5, 10\}$ and boundary knots placed at times 0 and 25. For the error terms we assumed $\varepsilon_i(t) \sim \mathcal N(0, \sigma^2)$, with $\sigma = 0.126$. The random effects were assumed to follow a multivariate normal distribution with mean zero and variance-covariance matrix $\bD$ given by:
\[
\bD =
\begin{bmatrix}
1.3343607 &  0.17546590 & 0.27199009 &  0.28257378\\
0.1754659 &  0.09889257 & 0.05086370 & -0.01836841\\
0.2719901 &  0.05086370 & 1.56264217 &  0.05125092\\
0.2825738 & -0.01836841 & 0.05125092 &  0.10579131
\end{bmatrix}.
\]

The fixed effects are, a nonlinear time effect modeled with natural cubic splines with three degrees of freedom, and the random effects, random intercepts, and nonlinear random slopes with the same spline specification as in the fixed effects. For each subject, we simulated longitudinal responses at time zero and then at 14 randomly selected time points coming from $\mbox{Unif}(0, 25)$. For the event time outcome, we simulated event times from the hazard model:
\[
h_i(t) = h_0(t) \exp \bigl \{\gamma_0 + \gamma_1 \mbox{treat}_i + \alpha \eta_i(t) \bigr \},
\]
where $h_0(t) = \phi t^{\phi - 1}$, with $\phi = 6.325$, $\gamma_0 = -20$, $\gamma_1 = -0.85$, $\mbox{treat}_i$ denotes a binary treatment indicator, and $\alpha = 0.145$. The censoring mechanism was fixed Type I right censoring, i.e., all event times greater than 25 were censored. The longitudinal measurements that were taken after the observed time $T_i = \min(T_i^*, C_i)$ (with $T_i^*$ denoting the true event time, and $C_i$ the censoring time) were removed. From the 300 subjects, 189 had an event, and we had an average of 10 ($\mbox{s.d.} = 2.2$) longitudinal measurements per subject.

We analyzed this dataset using four joint models with a different specification of the longitudinal and survival submodels. First, we fitted the true model from which we simulated the dataset. In the second model, we misspecified the longitudinal submodel by assuming linear subject-specific trends (i.e., in both the fixed- and random-effects parts we have an intercept and a linear time trend). In the third model, we misspecified the longitudinal submodel by fitting the mixed model for the outcome variable $y^* = \exp(y)$. In the final model, we specified the current slope/velocity as the functional form in the survival submodel, i.e., $\eta_i'(t) = \dd \eta_i(t) / \dd t$. All other aspects of the misspecified models were identical to those of the data-generating mechanism.

The results are presented in Section~2 of the supplementary material. From the table in Section~2.1 showing DIC, WAIC, and LPML, we observe that for this dataset, all criteria select the joint model with the current slope $\eta_i'(t)$ functional form, even though it is misspecified. The eCDF checks for the longitudinal outcome indicate that this metric captures only the misspecification due to the transformation (i.e., the misspecified longitudinal submodel with $y^*$ as the outcome). The checks for the mean function clearly identify the misspecification of the fixed- and random-effects parts assuming a linear time trend. Interestingly, we see that the simulated mean functions from the misspecified model with $y^*$ as the outcome are close to the mean function of the observed data, although some departure from a good fit is evident. The misspecification of the functional form in the survival submodel does not influence the mean function for the longitudinal part. The checks for the variance function show that misspecification of either the mean subject-specific structure or the transformation results in a misfit. We can make similar observations in the checks for the correlation structure of the repeated measurements, as with the mean function: the misspecification of the longitudinal submodel's mean structure has a large influence, and the transformation of the outcome results in only a mild difference between the simulated and actual data. The eCDF and probability integral transform checks for the event time submodel show a good fit for all misspecified models. Finally, the checks for the association structure between the two processes indicate that misspecification due to the transformation of the longitudinal outcome and the functional form in the survival model has the greatest impact in this part of the joint distribution.

%=====================================================

\section{Discussion} \label{Sec:Discussion}
In this paper, we have proposed a posterior predictive check framework for assessing the fit of a joint longitudinal-survival model. Joint models have different components, the main ones being the mean and covariance structure of the longitudinal submodel, the structure of the survival submodel, and the link between the longitudinal and event processes. We presented goodness-of-fit measures that were intended to target each of these components separately. The simulations and the data example illustrate the methods and suggest that they are effective at finding model misspecifications.

The goodness-of-fit metrics we introduced were based on empirical distribution functions, means, variances, semi-variograms, concordance statistics, Kaplan-Meier plots, and probability integral transformations. Alternative or additional metrics could be used. These could easily be incorporated into the proposed posterior predictive checks framework. Of the various goodness-of-fit plots we are suggesting, the eCDF plot is perhaps the least useful for detecting important misspecifications. It detected the fairly extreme transformation in the longitudinal outcome we used in the simulation study. Still, even there, the difference between the simulated lines and the actual data lines was not substantial. In the Bio-SHIFT data, the eCDF plot for eGFR showed a mild misspecification. This was primarily due to right skewness and outliers in that variable. As part of the analysis of those data, we repeated the analysis using the square root of eGFR and assuming the error terms followed a Student's t distribution. This analysis led to a mild improvement in the eCDF plot, but it had little effect on the other plots and did not alter any findings from the parameter estimates and credible intervals.

One part of the model we did not attempt to assess was how baseline covariates are included in both the longitudinal and the survival model. A simple approach to address this might be to bin the baseline covariates into, say, two or three levels and repeat the proposed set of goodness-of-fit checks separately for each bin. This approach could be useful for suggesting interactions or non-linear functional forms for continuous covariates or for assessing the proportional hazards assumption in the survival submodel. Alternatively, the concordance statistic we used for the association between the longitudinal and event-time outcomes can, in principle, be used for baseline covariates as well.

The mean, variance, and correlation metrics we have presented focused on continuous longitudinal outcomes. In the context of categorical longitudinal outcomes, the loess-based comparison between the observed and simulated data for the mean function may still be applicable; however, the variance and semi-variogram figures based on the loess residuals may be less appropriate. For such outcomes, other measures will be needed.

All proposed methodology is implemented in the freely available \textbf{JMbayes2} package. The vignette `Posterior Predictive Checks' (accessible via the URL: \url{https://drizopoulos.github.io/JMbayes2/articles/Posterior_Predictive_Checks.html}) illustrates the use of the package for implementing the various checks.

\section*{Funding}
This research was partially supported by US National Institutes of Health grant CA46592.

\section*{Supplementary Material}
Supplementary material with this paper are available at \url{https://www.drizopoulos.com/html/ppc_jms/}.

%=====================================================

\bibliographystyle{biom}
\bibliography{PPC_JMs.bib}

%=====================================================
\end{document}